# Relationship between chain reactions in a nuclear reactor and multifractality

V. V. Ryazanov

Institute for Nuclear Research, pr. Nauki, 47 Kiev, Ukraine, e-mail: vryazan19@gmail.com

A multifractal model of neutron evolution in a reactor is considered. For chain reactions, the dimension of the multifractal carrier, information and correlation dimensions, the entropy of the fractal set, the maximum and minimum values of the dimension, the multifractal spectrum function and other characteristics of multifractal neutron behavior are found. The use of geometric characteristics of a multifractal makes it possible to describe a stochastic system of hierarchically subordinate statistical ensembles, characterized by Cayley trees. A stationary distribution over hierarchical levels is established, which reduces to the Tsallis power law. Some possibilities for using fractal patterns in the theory of nuclear reactors are indicated.

1. Введение

Fractals with variable dimensions - multifractals [1, 2] are more common in nature than exact self-similarities and fractals with constant dimensions, which are rare under normal conditions. A multifractal is a non-uniform fractal object. Fractal sets, representing real physical formations and processes, have, as a rule, not one dimension, but a spectrum of dimensions. In addition to the geometric characteristics inherent in monofractals with one dimension, such sets also have statistical properties. The neutron fission chains in a chain reaction in a reactor are no exception.

In accordance with [1, 2], we define the fractal dimension. The space in which the fractal object is located is called the embedding space. This is an ordinary Euclidean space with dimension d. Let us cover the entire fractal object with d-dimensional "balls" of radius l. If this required at least N(l) balls, and for sufficiently small l, the value of N(l) changes with l according to a power law

$$N(l) \sim l^{-D},$$

then D is called the Hausdorff or fractal dimension of this object [1, 2].

A multifractal is characterized by its spectrum; its dimension varies depending on all values of the non-additivity parameter q, a measure of the multifractal's heterogeneity. The non-additivity parameter q appears in non-additive thermodynamics and in statistical distributions, for example, the Renyi and Tsallis distributions [3], indicating the degree of deviation from additivity. At q=1, thermodynamics becomes additive, and the Tsallis distribution becomes Gibbsian. For natural fractals, there is a minimum scale at which the main property of a fractal—self-similarity—disappears. For neutrons in a reactor, these are values on the order of the neutron free path $\lambda$. The self-similarity property is also violated at sufficiently large scales. For reactors, these are correlation lengths, the characteristic spatial scale of the cluster, corresponding to the critical size of the reactor near the critical point. How to carry out a fractal description on finite scales is discussed, for example, in [4].

The study of multifractality and the possibilities of its quantitative description is a new method for analyzing experimental data, applicable to a wide range of problems. In [5-7], multifractal analysis of complex signals is described, applicable, for example, to non-stationary reactor power. Using multifractal fluctuation analysis, time series in economics, medicine, and meteorology were studied. It can be used no less successfully in various problems of the theory of nuclear reactors.

The issues raised in this work have many other diverse aspects. Thus, the use of geometric characteristics of a multifractal makes it possible to describe a stochastic system consisting of hierarchically subordinate statistical ensembles and characterized by Cayley trees [6-8], as the evolution of neutrons in a reactor. In [6, 7] it is shown that the evolution of hierarchical structures reduces to the process of anomalous diffusion in ultrametric space, as a result of which a stationary distribution over hierarchical levels is established, which reduces to the Tsallis power law characteristic of non-additive systems [3, 9]. The Tsallis distribution is a special case of superstatistics [10], related to the distribution containing the lifetime [11], which was used to study processes in a nuclear reactor [12, 13].

The evolution of neutrons in the reactor follows fractal trajectories like Cayley trees. This is confirmed by the successful description of neutron processes in a reactor by random branching processes [14, 15]. Many of the results obtained for fractal structures can find application in the theory of nuclear reactors. In particular, the inhomogeneous Sierpinski triangle can serve as an effective model for describing chain reactions in a reactor.

Anomalous diffusion, as well as such characteristics as the time of first reaching the level, etc., on the Sierpinski triangle and branching structures are considered, for example, in [16]. An analogy can be drawn with statistical physics, in which the Ising model serves as an effective model. Reactor characteristics such as critical size, power, disturbance propagation speed, etc. are associated with fractal geometry. Therefore, taking into account fractal patterns when describing the behavior of neutrons in a reactor seems useful and necessary. At the same time, many well-known phenomena are examined from a different point of view, making it possible to discover some new features of already well-known things. Many results of the theory of fractals can find application in the theory of reactors.

The multifractal description turns out to be effective when applied to a wide variety of physical, chemical, and biological problems. It can be assumed that the study of such features of the behavior of a group of neutrons in a nuclear reactor will make it possible to understand in more detail some issues important for the safe operation of reactors: to clarify, for example, the dependence of the position of the critical point on the size of the system, the critical dimensions of the reactor, the speed of propagation of disturbances in the reactor, etc. In [17], the author, illustrating the phenomenon of percolation and percolation he describes, compares the spread of rumors in the percolation model with a chain reaction and writes that rumors are similar to the explosion of an atomic bomb. The similarity between the fractal geometry of Cayley trees and neutron multiplication processes in reactors can also be justified by the fact that both of these processes are associated with random branching processes, Galton-Watson processes [14, 15].

Neutron trajectories in the reactor represent real-life fractal structures (Cayley trees). Therefore, neutron processes in a reactor should be studied, in addition to other methods, using the methods of extensive science about fractals. There are many other self-similar fractal structures: a velocity field in a turbulent fluid flow, a lightning discharge in the air, the shape of clouds, the human circulatory system, the contours of a tree, etc. Note that now multifractal analysis is successfully used to describe the structural distribution of inhomogeneous star clusters in astrophysics, in the study of aggregation properties of cellular blood elements in biology, to characterize the main stages of the evolution of an ensemble of dislocations and fatigue fracture of materials in metal physics. Multifractal concepts are widely used in the theory of developed hydrodynamic turbulence, in the study of incommensurate structures and quasicrystals in solid state physics, in the theory of spin glasses and disordered systems, in quantum mechanics and elementary particle physics [2].

## 2. Multifractality of fission and neutron transfer processes in the reactor

To determine the distribution of the relative density of the number of neutrons, as in [1, 2], we divide the entire reactor region of size L into cubic cells with a side of size $\varepsilon_0 = L / (L / \varepsilon_0) = l_0 << L$ (L/$\varepsilon_0$ is the number of cells) and volume $\varepsilon_0^d$, where $l_0$ is of the order of the neutron mean free path $\lambda$. The neutron population, assuming that the initial neutron is one, consists of N neutrons distributed throughout the reactor volume. In the limit of an infinite multiplying medium N→∞. Following the procedure applied in [1, 2] to the inhomogeneous Cantor set, the inhomogeneous Sierpinski triangle and other fractal objects, we assume that at the n-th step, for the n-th generation of neutrons, the number of cells changes as $(L/\varepsilon_0)^n$, and the cell size becomes equal to $\varepsilon_n = L/(L/\varepsilon_0)^n$.

This exponential increase in the number of cells is due to the fact that in Cayley trees corresponding to division processes, the number of nodes and event possibilities increases exponentially. Let us denote the cells by the index I = 0, 1, 2, …, N - 1. The distribution of the neutron population throughout the reactor volume is characterized by a set of numbers $n_i$, indicating how many neutrons are in the i-th cell. Magnitude

$$p_i = n_i/N = p_i(\varepsilon) = \lim_{N\to\infty} n_i(\varepsilon)/N \tag{1}$$

represents the probability that a randomly selected neutron is located in the i-th cell. In a real situation, we always have a finite (albeit large) number of neutrons N. The cell size is equal to $\varepsilon_0 = L/(L/\varepsilon_0) = l_0$ only at the zero step, before the start of the division process corresponding to the procedure for constructing a fractal set - the Cayley tree. The cell size in the n-th generation of neutrons, as indicated above, is equal to $\varepsilon_n = L/(L/\varepsilon_0)^n$ and tends to zero as n→∞ (although in real reactors the values of n are finite, but can be very large).

Let us first consider a simple model without distinguishing prompt and delayed neutrons. We use the approximation from the theory of branching processes [6, 7], in which one nuclear fission produces either 0 neutrons with probability $\pi_0$ or 2 neutrons with probability $\pi_2$, $\pi_0 + \pi_2 = 1$. With probability one neutron will collide with a fuel nucleus and split it ($\lambda_f = v\Sigma_f$, v - neutron speed, $\Sigma_f$ - macroscopic fission cross section, $\lambda_f$ - fission intensity, the absorption intensity $\lambda_c$ is determined similarly. The value p is calculated with known values of the cross sections $\sigma_f$, $\sigma_c$. The relations $k_{eff} = pv$ are valid, where $k_{eff}$ is the effective neutron multiplication factor, v is the average number of neutrons produced during one fission of a fuel nucleus. Note that the values and $p_i(\varepsilon)$ do not coincide. Thus, after the first division, the probability that in a cell with size $\varepsilon_2 = L/(L/\varepsilon_0)^2$ there are 0, 1 and 2 neutrons are equal to: $p_0(\varepsilon) = p\pi_0$, $p_1(\varepsilon) = 1-p$, $p_2(\varepsilon) = p\pi_2$. Then these probabilities become more complex. The sum of the probabilities of all possible events after one fission

$$\Phi = p(\pi_0 + \pi_2) + 1 - p; \quad \Phi = 1; \quad (\pi_0 + \pi_2 = 1). \tag{2}$$

In relations (1) and (2), the probability normalization condition $\sum_{i=1}^{N} p_i(\varepsilon) = 1$, $\Phi = 1$ is satisfied. The generalized partition function is characterized by the exponent q, $-\infty < q < \infty$ [2]

$$Z(q,\varepsilon) = \sum_{i=1}^{N} p_i^q(\varepsilon). \tag{3}$$

The spectrum of generalized fractal dimensions is defined in [2] by the relation

$$D_q = \frac{\tau(q)}{q-1}, \tag{4}$$

where

$$\tau(q) = \lim_{\varepsilon \to 0} \frac{\ln Z(q,\varepsilon)}{\ln \varepsilon}. \tag{5}$$

When $D_q = D = const$ we obtain the usual regular monofractal.

To characterize the division process, we use the algorithm described in [2], when the process is binomial and the generalized statistical sum, constructed in accordance with definition (3), for our multifractal has the form

$$Z(q,\varepsilon) = (p_0^q + p_1^q + p_2^q)^n = \Phi_q^n, \quad \Phi_q = p^q(\pi_0^q + \pi_2^q) + (1-p)^q. \tag{6}$$

Two potentially produced neutrons can lead to further events with a sum of probabilities $\Phi^2$. In the second generation the probabilities are $\Phi^3$, in the third – $\Phi^4$, in the n-th – $\Phi^n$. From formulas (3), (5) and (6) we obtain that for $\varepsilon_n \to 0$, n→∞,

$$Z(q,\varepsilon) = \sum_{i=1}^{N} p_i^q(\varepsilon) \approx \varepsilon^{\tau(q)}. \tag{7}$$

From relations (4) - (7) for $Z(q,\varepsilon) = \Phi_q^n$, $\Phi_{q=1} = 1$ and $\varepsilon_n = L/(L/\varepsilon_0)^n$ we obtain for large n:

$$\tau(q) = (q-1)D_q = \frac{\ln \Phi_q}{\ln(L/\varepsilon_n) - \ln(L/\varepsilon_0)} \simeq \frac{\ln \Phi_q}{\ln(L/\varepsilon_0)}. \tag{8}$$

Putting q = 0 in expression (8), we find the dimension

$$D_0 = -\frac{\ln 3}{\ln(l_0/L)}. \tag{9}$$

This is the so-called dimension of the multifractal carrier [2]. At q = 1

$$D_1 = -\frac{\ln S}{\ln(l_0/L)}, \quad -S = \sum_{i=1}^{N} p_i \ln p_i = p\pi_0 \ln(p\pi_0) + p\pi_2 \ln(p\pi_2) + (1-p)\ln(1-p), \tag{10}$$

where S is the entropy of the partition of the measure on the set L, the entropy of the fractal set. At q = 2

$$D_2 = \frac{\ln[p^2(\pi_0^2 + \pi_2^2) + (1-p)^2]}{\ln(l_0/L)}. \tag{11}$$

Having defined the pair correlation integral $I(l) = \lim_{N \to \infty} \frac{1}{N^2} \sum_{n,m}^{N} \theta(l - |r_n - r_m|)$, where the summation is carried out over all pairs of points of the fractal set with radius vectors $r_n$ and $r_m$, $\theta(x)=1$, if $x \geq 0$, $\theta(x)=0$, if $x < 0$, we obtain that $I(l) \approx \sum_{i=1}^{N} p_i^2 \approx (\frac{l}{L})^{D_2}$.

If we introduce the probability density $\rho(r) = \frac{1}{N} \sum_i \delta(r - r_i)$, $\int_L \rho(r)d^d r = 1$, then the pair correlation function, which is the probability density for two arbitrary points of the set to be at a distance r relative to each other, is equal to $C(r) = \int_L \rho(r')\rho(r'+r)d^d r'$. The correlation function C(r) is characterized by power-law behavior with distance r, i.e. $C(r) \approx 1/r^\beta$, $\beta = d - D_2$, where d is the Euclidean dimension of space, equal to 3 in our case. The Fourier component C(k), depending on the wave vector k, also changes according to a power law $C(k) = 1/k^{D_2}$.

The maximum value of the dimension is

$$D_{max} = D_- = \frac{\ln(p\pi_0)}{\ln(l_0/L)}, \tag{12}$$

its value $D_q$ reaches at q→-∞. Minimum dimension value

$$D_{min} = D_- = \frac{\ln(1-p)}{\ln(l_0/L)} \tag{13}$$

is achieved at q→∞. In formulas (12) and (13) we assumed that, in accordance with the fact that $k_{ef} = p\bar{V}$, with $k_{ef} \simeq 1$, $\bar{V} = 2,4$, $p \simeq 1/\bar{V} \simeq 0,4$, and put $\pi_2 = 0,8$, $\pi_0 = 0,2$.

Together with the generalized dimensions $D_q$, the multifractal spectrum function (or spectrum of multifractal singularities) $f(\alpha)$ is used to characterize a multifractal set [1, 2]. The set of probabilities $p_i$ (1), showing the relative population of cells ε that cover the set, for self-similar sets has a power-law character depending on the size of the cell ε:

$$p_i(\varepsilon) \simeq \varepsilon^{\alpha_i}, \tag{14}$$

where $\alpha_i$ is the exponent. α values fill the interval ($\alpha_{min}$, $\alpha_{max}$),

$$p_{min} \simeq \varepsilon^{\alpha_{max}}, \quad p_{max} \simeq \varepsilon^{\alpha_{min}}; \quad \frac{d\tau}{dq}\bigg|_{q \to +} = D_- = \alpha_{min}; \quad \frac{d\tau}{dq}\bigg|_{q \to -} = D_- = \alpha_{max}. \tag{15}$$

Knowing $D_q$, one can find the dependence $\alpha(q)$ from the equation

$$\alpha(q) = \frac{d\tau}{dq}[(q-1)D_q]. \qquad (16)$$

For $\Phi_q$ (2), expression (16) takes the form

$$\alpha(q) = \frac{1}{\ln(l_0/L)}\frac{1}{\Phi_q}\frac{d\Phi_q}{dq}; \quad \frac{d\Phi_q}{dq} = (\ln p)p^q(\pi_0^q + \pi_2^q) + p^q(\pi_0^q \ln\pi_0 + \pi_2^q \ln\pi_2) + (1-p)^q \ln(1-p)$$

From the expression

$$\tau(q) = q\alpha(q) - f(\alpha(q)); \quad f(\alpha(q)) = \frac{q}{\ln(l_0/L)\Phi_q}\frac{d\Phi_q}{dq} - \frac{1}{\ln(l_0/L)}\ln\Phi_q \qquad (17)$$

the dependence $f(\alpha(\ ))$ is determined (in parametric form), i.e. from formula (16) we find $q(\alpha)$ and substitute it into expression (17). The variables $\{q, \tau(q)\}$ are related to the variables $\{\alpha, f(\alpha)\}$ by the Legendre transform $\alpha = \frac{d\tau}{dq}; \quad f(\alpha) = q\frac{d\tau}{dq} - \tau$. Inverse Legendre transform

$$q = \frac{df}{d\alpha}; \quad \tau(q) = \alpha\frac{df}{d\alpha} - f; \quad \frac{d^2f}{d\alpha^2} = (\frac{d^2\tau}{dq^2})^{-1}.$$

At the point $\alpha_0 = \alpha(0)$ the function $f(\alpha)$ (convex everywhere) has a maximum. For $\Phi_q$ (2)

$$\alpha_0 = \alpha(q=0) = \frac{q}{\ln(l_0/L)}\frac{1}{3}[2\ln p + \ln\pi_0 + \ln\pi_2 + \ln(1-p)].$$

The fractal dimension of the measure support is equal to $f(\alpha_0) = D_0$. The function $f(\alpha)$ near its maximum can be approximated by a parabola. Because

$$\tau'(0) = \alpha_0; \quad f''(\alpha) = 1/\tau''(0); \quad \tau''(0) = 2(D_0 - \alpha_0) - D''_{q=0}, \text{ that}$$

$$f(\alpha) \simeq D_0 - \frac{(\alpha - \alpha_0)^2}{2[2(\alpha_0 - D_0) + D''_{q=0}]}.$$

For $\Phi_q$ from expression (2)

$$D_q = \frac{1}{\ln(l_0/L)\Phi_q}\ln\Phi_q; \quad \frac{dD_q}{dq} = \frac{1}{\ln(l_0/L)}(-\frac{\ln\Phi_q}{(q-1)^2} + \frac{1}{(q-1)\Phi_q}\frac{d\Phi_q}{dq});$$

$$\frac{d^2D_q}{dq^2} = \frac{1}{\ln(l_0/L)}\{\frac{2\ln\Phi_q}{(q-1)^3} - \frac{2}{(q-1)^2\Phi_q}\frac{d\Phi_q}{dq} + \frac{1}{(q-1)}[\frac{1}{\Phi_q}\frac{d^2\Phi_q}{dq^2} - (\frac{1}{\Phi_q}\frac{d\Phi_q}{dq})^2]\};$$

$$\frac{d^2\Phi_q}{dq^2} = (\ln p)^2 p^q(\pi_0^q + \pi_2^q) + 2(\ln p)p^q(\pi_0^q \ln\pi_0 + \pi_2^q \ln\pi_2) + p^q[\pi_0^q(\ln\pi_0)^2 + \pi_2^q(\ln\pi_2)^2] + (1-p)^q(\ln(1-p))^2;$$

$$\frac{d^2D_q}{dq^2}\bigg|_{q=0} = D''_{q=0} = \frac{1}{\ln(l_0/L)}\{-2\ln 3 - \frac{2}{3}[2\ln p + \ln\pi_0 + \ln\pi_2 + \ln(1-p)] + [\frac{1}{3}(2\ln p + \ln\pi_0 + \ln\pi_2 + \ln(1-p))]^2 -$$

$$-\frac{1}{3}[2(\ln p)^2 + 2\ln p(\ln\pi_0 + \ln\pi_2) + (\ln\pi_0)^2 + (\ln\pi_2)^2 + (\ln(1-p))^2]\}$$

When $q=1$, $\alpha(1)=D_1=f(\alpha(1))$. When $q=2$, $f(\alpha(2))=2\alpha(2)-D_2$.

From expression (14) $\alpha_i \simeq \ln p_i / \ln(l_0/L)$. The distribution of $\alpha$ values for a multifractal is determined by the relation $n(\alpha) \simeq (\frac{l_0}{L})^{-f(\alpha)} = e^{-f(\alpha)\ln(l_0/L)}$.

When approximating the function $f(\alpha)$ near its maximum $\alpha_0$ by a parabola

$$f(\alpha) \simeq D_0 - \eta(\alpha - \alpha_0)^2; \quad \eta = f''(\alpha_0)/2; \quad n(\alpha) \sim \exp\{-\ln(L/l_0)(\alpha - \alpha_0)^2\}.$$

Since $\alpha_i \simeq \ln p_i / \ln(l_0/L)$ then the probability distribution function $p_i$ is equal to

$$P(p) \sim \exp\{-\eta \ln(L/l_0)(\frac{\ln p}{\ln(L/l_0)} + \alpha_0)^2\}. \tag{18}$$

This is a log-normal distribution. Probabilities $p_i$ characterize the location of neutrons, their density and reactor power (the role of the coordinate is played by the index i). Relation (18) shows that the neutron density in the critical region is distributed log-normally over the reactor volume. The distribution of thermal neutron flux density along the radius of the core of a cylindrical reactor is described by a Bessel function of the first kind of zero order. Within the relatively thin boundary layer, the distribution curve rises slightly. In certain approximations, this behavior of the function can be approximated by the right wing of the log-normal distribution. Note that approximation (18) is very rough; it does not take into account the geometry of the reactor. Therefore, it is more correct to consider expression (18) as an approximation of the spectral function $f(\alpha)$ by a parabola. One more interpretation of relation (18) can be given. The quantity $\alpha$ (16) is proportional to the quantity $\sum_{i=1}^{N} p_i^q \ln p_i$, which at $q=1$ coincides with entropy (10). And expression (18) as $q \to 1$ turns out to be related to the entropy distribution.

From the model function $\Phi_q$ from expression (2) we move on to a more realistic situation, when the relation $k_{ef} = p\bar{\nu}$, are valid, the probabilities of the occurrence of I = 0, 1, ..., 7 secondary neutrons during the fission of $^{235}U$ by thermal neutrons are equal to: $\pi_0 = 0,0333$, $\pi_1 = 0,1745$, $\pi_2 = 0,3349$, $\pi_3 = 0,3028$, $\pi_4 = 0,1231$, $\pi_5 = 0,0281$, $\pi_6 = 0,0032$, $\pi_7 = 0,0001$ [15]. Since fission produces two fragments (precursor nuclei), each of which, after some time T (about $10^3$ times longer than the neutron lifetime) can emit a delayed neutron, we introduce the probabilities $r_0(T)$, $r_1(T)$, $r_2(T)$ that as a result of one fission I = 0, 1, 2 delayed neutrons are produced, $\sum_{i=0}^{2} r_i = 1$.

The T value is the half-life of the precursor nuclei. There are 6 groups of delayed neutrons depending on the delay time. In this work, kinetics is not considered and the dependence of $r_i(T)$ on the period T is not taken into account. The value $\Phi_q$ from expression (2) is replaced by a more complex function of the form

$$\Phi_q = (1-\beta)^q [p^q \sum_{i=0}^{7} \pi_i^q + (1-p)^q] + \beta^q [p^q \sum_{i=0}^{2} r_i^q(T) + (1-p)^q], \tag{19}$$

where for $^{235}U$, $\beta = 0.0064$. Instead of formulas (8) - (10) we get:

$$D_0 = \ln \Phi_{q=0} / \ln(l/L) \simeq 0557, \quad \Phi_{q=0} = 13, \quad l_0/L \approx 10^{-2}.$$

$$-S = (1-\beta)[p \sum_{i=0}^{7} \pi_i \ln(p\pi_i) + \ln(1-\beta)] + \beta[p \sum_{i=0}^{2} r_i(T) \ln(pr_i) + \ln(\beta)] + (1-p)\ln(1-p), \quad D_1 = 0.294,$$

$$D_2 = \frac{1}{\ln(l_0/L)} \ln\{(1-\beta)^2 [p^2 \sum_{i=0}^{7} \pi_i^2 + (1-p)^2] + \beta^2 [p^2 \sum_{i=0}^{2} r_i^2(T) + (1-p)^2]\}, \quad D_2 = 0.211.$$

The values of $D_{max} = D_- = \alpha_{max}$ and $D_{min} = D = \alpha_{min}$ (15) depend on the minimum and maximum values of the products $(1-\beta)p\pi_i$, $i=0,...,7$, $\beta r_i$, $i=0,1,2$, $(1-\beta)(1-p)$, $\beta(1-p)$, where values $r_i$, $i=0, 1, 2$, not defined. If you set $r_2 = r_0 = 0.1$, $r_1 = 0.8$, then the minimum value is $(1-\beta)p\pi_7$, and $D_- = \frac{1}{\ln(l_0/L)} \ln[(1-\beta)p\pi_7] \approx 2.1$. The maximum combination in this case will be $(1-\beta)(1-p)$, and

$$D = \frac{1}{\ln(l_0/L)} \ln[(1-\beta)(1-p)] \approx 0.125.$$

From relations (3) - (8) and (19) we construct a spectrum of generalized dimensions (Fig.

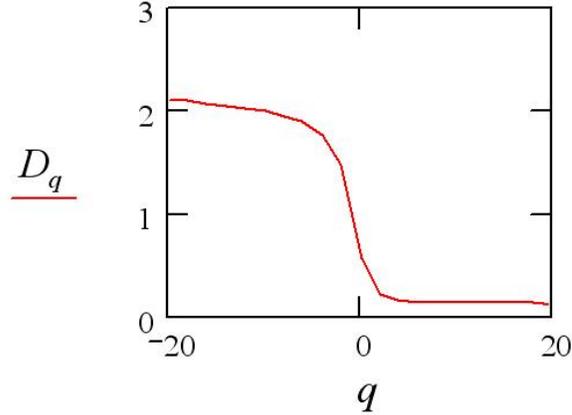

1).
Fig. 1. Spectrum of generalized dimensions for fission chains in nuclear reactors, taking into account delayed neutrons.

Just as above, the values are determined from formula (19) $\frac{d\Phi_q}{dq}\Big|_{q=0}$, $\frac{d^2\Phi_q}{dq^2}\Big|_{q=0}$ and

$$\alpha_0 = \alpha(q=0) = \frac{1}{\ln(l_0/L)}\frac{1}{13}[9\ln(1-\beta)+8\ln p+\sum_{i=0}^{7}\ln\pi_i+\ln(1-p)+4\ln\beta+3\ln p+\sum_{i=0}^{2}\ln r_i+\ln(1-p)].$$

In the expression for η from formula (18), the value is worth

$$D''|_{q=o} = \frac{1}{\ln(l_0/L)}[-2\ln 13 - \frac{2}{13}\frac{d\Phi_q}{dq}\Big|_{q=o} + (\frac{1}{13}\frac{d\Phi_q}{dq}\Big|_{q=o})^2 - \frac{1}{13}\frac{d^2\Phi_q}{dq^2}\Big|_{q=o}],$$

$$\frac{d\Phi_q}{dq}\Big|_{q=0} = 9\ln(1-\beta)+8\ln p+\sum_{i=0}^{7}\ln\pi_i+2\ln(1-p)+4\ln\beta+3\ln p+\sum_{i=0}^{2}\ln r_i,$$

$$\frac{d^2\Phi_q}{dq^2}\Big|_{q=0} = 9[\ln(1-\beta)]^2+2\ln(1-\beta)[8\ln p+\sum_{i=0}^{7}\ln\pi_i+\ln(1-p)]+4(\ln\beta)^2+11(\ln p)^2 +$$

$$+2\ln p(\sum_{i=0}^{7}\ln\pi_i+\sum_{i=0}^{2}\ln r_i)+\sum_{i=0}^{7}(\ln\pi_i)^2+2(\ln(1-p))^2+2\ln\beta(3\ln p+\sum_{i=0}^{2}\ln r_i+\ln(1-p))+\sum_{i=0}^{2}(\ln r_i)^2$$

In the Log-Normal Distribution of the species (18), combinations of values appear as probability p: $(1-\beta)p\pi_i$, $i=0,...,7$, $\beta pr_i$, $i=0,1,2$, $(1-\beta)(1-p)$, $\beta(1-p)$.

If we construct the dependence of the multifractal spectrum f(α) from formula (16), using the approach of [1, 18], with a measure of the multiplicative population, we obtain the dependence as in Fig. 2

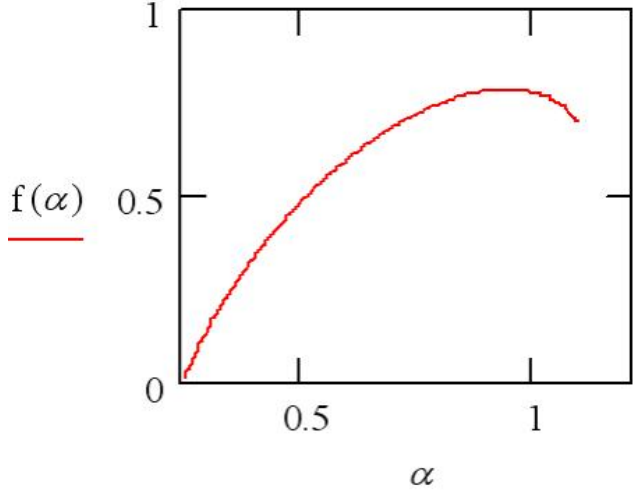

Fig. 2. Multifractal spectrum function for fission chains in nuclear reactors taking into account delayed neutrons. The function f(α) for the inhomogeneous Sierpinski triangle has a similar form [2].

### 3. Conclusion

The number of practical applications of fractal patterns in the theory of nuclear reactors can be very large. To illustrate, let us point out the result obtained in [5] for the size of the critical region in which the values of the percolation threshold, the critical point $k_{ef}$, can be located. The width of this region is $\delta = (l/L)^{1/\nu}$, where $\nu$ is the index of the correlation radius corresponding to the critical size of the reactor $R_{cr} = \pi M(k_{ef}-1)^{1/2}$, $\nu = 1/2$, where $M = (L_T + \tau_T)^{1/2}$ is the neutron migration length [19], $L_T$ is the neutron diffusion length at temperature T, $\tau_T$ is the age of thermal neutrons [19], l is a value on the order of the lattice size, on the order of the neutron mean free path in the reactor, $L\;\;R_{cr}$, $L \simeq R_{cr}$. For a reactor, the value δ is rewritten in the form $\delta = (\pi M / L\sqrt{\bar{\nu}})^2$, where $\bar{\nu}$ is the mathematical expectation of the number of secondary neutrons in one fission event. At the critical point itself, the quantities L and $R_{cr}$ tend to infinity, the quantity δ tends to zero, and the critical region turns to a point. In [17], the dependences of the value of δ on the total number of nodes N and the number of neutrons are also indicated. The half-width of the distribution of critical point values is equal to $\Delta_N = 2(2\ln 2)\delta(N)$, $\delta(N) = C/N^{1/\nu d}$, $(1/\nu d) = 2/3$, C is a numerical coefficient. But for systems with a not very large number of neutrons, for example, a critical assembly, a reactor during startup, etc., the value of N can be relatively small, and the critical point can, with a certain probability, take on different values in a finite interval. Fluctuations of the critical point occur. In [17], the probability was determined that the percolation threshold, which corresponds to the critical point of the reactor, differs from the average value $k_{ef}=1$ by a value lying in the interval δ. At large N, the resulting distribution function of the percolation thresholds turns into a sharp peak. All values of the percolation thresholds, except one, have zero probability; the percolation threshold turns from a random value into a reliable value. The relations obtained in [17] make it possible to clarify the behavior of modes of approaching the critical point during reactor startup, which are important for the safety of nuclear power plants.

Another result from [17] concerns the fraction of nodes belonging to the skeleton of an infinite cluster, points that have at least two paths leaving them in different directions. For values of critical indices valid for the reactor, the proportion of nodes belonging to the skeleton of an infinite cluster at the critical point coincides in order of magnitude with the proportion of nodes of an infinite cluster. This corresponds to the situation when the fraction of dead ends of the cluster, neutron absorption points, at the critical point coincides in order of magnitude with the total number of nodes belonging to the infinite cluster, although for many other fractal systems, for example ferromagnets, the main "mass" of the infinite cluster is concentrated in the dead ends.

The kinetics and processes of transfer in fractal reactor structures are considered in [7, 8, 20, 21]. The problem arises of comparing the results obtained from equations in fractional derivatives, used to describe non-stationary processes on fractal structures [20, 21], with the results obtained from standard equations in Euclidean space. It was noted in [22] that in many similar situations the difference in results turns out to be insignificant.